\documentclass[10pt,letterpaper,twocolumn]{article} 
\usepackage{ol2}
\usepackage{hyperref}
\usepackage{amsmath}
\usepackage{amsfonts}
\usepackage{amssymb}
\usepackage{bbm}
\usepackage{widetext}
\usepackage{flushend}
\usepackage{cuted}
\newcommand{\TK}[1]{\textcolor{black}{ #1}}
\usepackage{epstopdf}
\usepackage{color}

\newcommand{\ket}[1]{\left| {#1} \right.\rangle}
\newcommand{\ex}[2]{\langle {#1}| {#2} \rangle}

\newcommand{\tr}{\mbox{Tr}}

\begin{document}

\twocolumn[ 

\title{Measuring the non-separability of classically entangled vector vortex beams}

\author{Melanie McLaren,$^{1}$ Thomas Konrad$^{2}$ and Andrew Forbes$^{1,3,*}$}
\address{{$^1$University of the Witwatersrand, Private Bag 3, Johannesburg 2050, South Africa}\\
{$^2$University of Kwazulu-Natal, Private Bag X54001, Durban 4000, South Africa}\\
{$^3$CSIR National Laser Centre, P.O. Box 395, Pretoria 0001, South Africa}\\
{$^*$Corresponding author: aforbes1@csir.co.za}}

\begin{abstract}
Given the \TK{multitude} of applications of vector vortex beams there is a need for robust tools to measure them.  Here we exploit the non-separability of such beams, akin to entanglement of quantum states, to apply tools traditionally associated with quantum measurements to these classical fields.  We apply three measures of non-separability: a Bell inequality, a concurrence, and an entanglement entropy to define the ``vectorness'' of such beams.  In addition to providing novel tools for the analysis of vector beams, we also introduce the concept of classical entanglement to explain why these tools are appropriate in the first place.
\end{abstract}
]
\begin{figure}[t]
\centerline{\includegraphics[width=8.3cm]{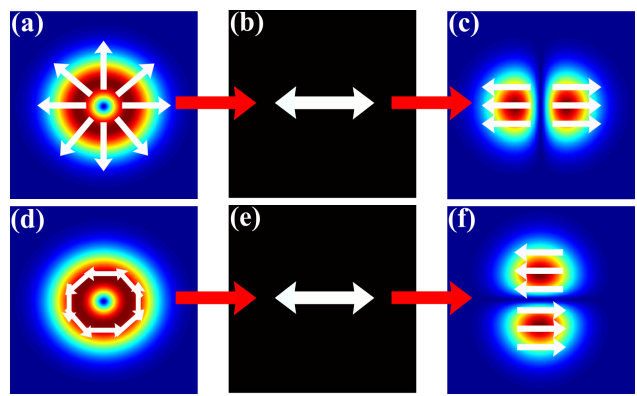}}
\caption{Schematic showing the spatial dependency of a vector beam on the polarization state. (a) A radially polarized vector beam incident on (b) a polarizer orientated to transmit horizontally polarized light produces (c) two petals orientated along the horizontal axis. (d) An azimuthally polarized vector beam incident on (e) a polarizer orientated to transmit horizontally polarized light produces (f) two petals orientated along the vertical axis.}
\label{CVbeams}
\end{figure}

\noindent 
Light beams with spatially inhomogeneous states of polarization, referred to as vector beams, have recently received increased interest in a variety of fields \cite{Zhan2009}. In particular, cylindrically symmetric vector (CV) beams have the ability to produce tighter focal spots with strong field gradients \cite{Novotny2001, Zhan2004}, finding applications in microscopy \cite{Abouraddy2006, Li2012}, interferometry \cite{Zhan2009} and optical tweezing \cite{Roxworthy2010}. CV modes have been observed in laser resonators \cite{Pohl1972,  Enderli2009}, optical fibers \cite{Grosjean2002, Zheng2010} and more recently generated by liquid crystal displays \cite{Han2011}, interferometric techniques \cite{Maurer2007} and q-plates \cite{Cardano2012}.  More general Higher-Order Poincar\'e sphere beams are the natural modes of many fibers \cite{Milione2011}.  A key characteristic of such vector fields is the coupling between the polarization and the spatial mode: in contrast to scalar fields, these degrees of freedom are non-separable, as depicted graphically in Fig.~\ref{CVbeams}.

Despite the coupling of the polarization and spatial modes, the existing methods of measuring vector beams do so by treating these degrees of freedom independently. For example, there has been a great deal of work in determining the spatial mode content of a beam, e.g., modal interference \cite{Nicholson2008}, phase-retrieval algorithms \cite{Shapira2005, Lu2013} and modal decomposition by digital holograms \cite{Flamm2012, Litvin2012}. Meanwhile the state of polarization of a beam is commonly measured using Stokes polarimetry where, at each point of the beam, the polarization orientation and ellipticity can be calculated \cite{Stokes1852}. This versatile tool has been used for the real-time monitoring of optical wavefronts during propagation \cite{Dudley2014} and for studying topological structures of polarization in vector vortex beams \cite{Cardano2012}.  But yet no measure exists for the overall ``vectorness'' of the field.

Here we employed measurement techniques more commonly associated with quantum entanglement experiments to determine the degree to which a vector beam is non-separable in spatial mode and polarization, in other words, the degree of vectorness of the field.  Our hypothesis is that since non-separablity is not unique to quantum mechanics, many of the tools for measuring this must be applicable to vector beams too.  We employ ubiquitous quantum tools to differentiate between scalar and vector beams: a Bell inequality measurement \cite{Aspect1981}, a concurrence measurement \cite{Jack2009} and an entanglement entropy measurement \cite{Wootters2001}.  We show that the former indicates if an unknown field is vector in nature, while the two other measurements allow the degree to which the field is vectorial to be measured, with a range from 0 (fully scalar) to 1 (fully vector).  In this way we offer new tools to determine the degree of ``vectorness'' of an unknown field.  We also discuss the implications of these findings for mimicking quantum processes with so-called classically entangled light.

We first derive a measure for the degree of vectorness of coherent paraxial beams and then for the more general case of incoherent mixtures of paraxial beams.
To begin, let us consider a paraxial  vector beam with frequency $\omega$ propagating along the z-axis, as represented by a complex-valued electric field $\mathbf{E}=E_0 e^{i\omega t} \mathbf{\Psi}$ with unit-amplitude complex vector field 
\begin{equation}
\mathbf{\Psi}(r,\phi,z) = \sqrt{a} u_{R}(r,\phi,z) \hat{\mathbf{e}}_R + \sqrt{(1-a)} u_{L}(r,\phi,z)\hat{\mathbf{e}}_L,
\label{eq:GeneralCV}
\end{equation} 
where $a$ determines the relative weighting of fields $u_{R}$ and $u_{L}$, which are normalized (i.e., $\int |u_{1,2}|^2 r dr d\phi =1$) and specify the spatial dependence of the right-handed and left-handed circular polarization components associated with the  canonical basis vectors $\hat{\mathbf{e}}_R$ and $\hat{\mathbf{e}}_L$. 

We can rewrite Eq.~(\ref{eq:GeneralCV}) in bra-ket notation \cite{Goyal2013} as
\begin{equation}
\left|\Psi \right\rangle = \sqrt{a} \left|u_{R}\right\rangle \otimes \left|R\right\rangle + \sqrt{(1-a)} \left|u_{L}\right\rangle\otimes \left|L\right\rangle,
\label{eq:QuantumCV}
\end{equation}
where  the kets $\ket{u_{R}}, \ket{u_{L}}$ are unit vectors in an infinite-dimensional Hilbert space $\mathcal{H}_\infty$ representing the complex spatial fields on a transversal plane (the corresponding parameter $z$ is omitted) and  $\left|R\right\rangle,\left|L\right\rangle \in\mathcal{H}_2$ stand for the  right-handed  and the left-handed circular polarization vectors, respectively. The symbol $\otimes$ denotes the tensor product between the vectors.
In quantum mechanics a state $\ket{\Psi}$ of the form in Eq.~(\ref{eq:QuantumCV}) is called non-separable or entangled, if it cannot be written as a product of any two vectors $\ket{u}\in\mathcal{H}_\infty$ and $\ket{P}\in \mathcal{H}_2$, i.e. $\ket{\Psi}\not=\ket{u}\otimes\ket{P}$. The non-separability of the state $\ket{\Psi}$ thus exactly matches the definition of a vector beam as a beam with  varying polarization over a transversal plane, i.e.,    
$\mathbf{E}$ can not be written as a product of a scalar field and a polarization vector,  which implies the non-separability of the state vector $\ket{\Psi}$ and hence, formally,  its entanglement. Therefore, measures of entanglement for quantum systems can be employed to measure the degree of vectorness of a vector beam. For example, when $a=1/2$ and the two modes $\ket{u_R}$ and $\ket{u_L}$ are orthogonal,  the field is purely vector (a maximally entangled state), whereas when $a=0$ or $1$, or the modes are the same, the field is purely scalar (a product state).   

What is the best suited entanglement measure in order to define vectorness? For pure states of bipartite quantum systems there is a fundamental entanglement measure on which most operational measures are based, the entanglement entropy \cite{Wootters2001}. Consider a two-partite state $\ket{\Psi}\in \mathcal{H}_\infty \otimes\mathcal{H}_2$ as used to describe the state of a paraxial beam. The entanglement entropy is then given by the von Neumann entropy of the reduced density matrix of one of the subsystems, for instance the polarization (P), which is obtained by tracing over the spatial degree of freedom (S):
\begin{equation}
E(\ket{\Psi})= -\tr[\rho_P\log(\rho_P)]\quad \mbox{with}\quad  \rho_P= \tr_S\left[\left|\Psi\right\rangle \left\langle\Psi\right|\right] 
\label{entanglemententropy}
\end{equation}
In order to understand the physical meaning of these operations in our optical context,  we first note that the density matrix $\rho$ of $\left|\Psi\right\rangle \left\langle\Psi\right|$ expressed with respect to the polarization basis $\ket{R}, \ket{L}$,
\begin{equation}
\rho= \begin{pmatrix}
  a\left|u_R\right\rangle \left\langle u_R \right| & \sqrt{a(1-a)}  \left|u_R\right\rangle \left\langle u_L \right|\\
  \sqrt{a(1-a)}  \left|u_R\right\rangle \left\langle u_L \right| & (1-a)\left|u_L\right\rangle \left\langle u_L \right|
  \label{BCP}
 \end{pmatrix}\,,
\end{equation}
corresponds to the beam coherence-polarisation matrix \cite{Gori1998}, which represents the state of light of vector beams. 
\TK{Averaging (tracing) $\rho$ over the spatial degree of freedom with an arbitrary set of orthonormal basis modes $\ket{b_i}$,} we  obtain a matrix  that resembles a polarization (or coherency) matrix as usually defined for scalar beams:
\begin{align}
\rho_P &= 
\sum_i \left( \begin{smallmatrix}
  a \ex{b_i}{u_R}\ex{u_R}{ b_i}  & \sqrt{a(1-a)}  \ex{b_i}{u_R}\ex{u_L}{ b_i} \\
  \sqrt{a(1-a)}  \ex{b_i}{u_L}\ex{u_R}{ b_i} & (1-a)\ex{b_i}{u_L}\ex{u_L}{ b_i} 
 \end{smallmatrix}\right) \, \label{pmatrix1}\\  
&= 
\begin{pmatrix}
  a  & \sqrt{a(1-a)}  \left\langle u_L | u_R \right\rangle\\
  \sqrt{a(1-a)}  \left\langle u_R | u_L \right\rangle & (1-a)
 \end{pmatrix}\,.
 \label{pmatrix}
\end{align}
However, the vector-beam equivalent of the polarization matrix for scalar beams can be argued to be the local polarization matrix $\tilde{\rho}(r,\phi, z)$= $\bigl(\begin{smallmatrix}
|u_r|^2 & u_Ru_L^*\\u_Lu_R^*& |u_r|^2
\end{smallmatrix} \bigr)$, where each matrix element depends on the spatial coordinates $r, \phi, z$, and defines in this way the polarization properties of the beam in each point in space \cite{Gori1998}. On the other hand, the reduced density matrix $\rho_P$  determines the {\sl average} polarization of the vector beam. It can be detected, for example, by measuring the components $r_i=\tr[\sigma_i \rho_P]$\footnote{Here $i=1,2,3$ and the Pauli operators $\sigma_i$ are given by 
\TK{$\sigma_1~=~\left|H \right\rangle \left\langle H \right |- \left|V \right\rangle  \left\langle V \right|$}, $\sigma_2=\left|H+V \right\rangle \left\langle H+V \right| - \left|H-V \right\rangle \left\langle H-V \right|$ and $\sigma_3=\left| R \right\rangle \left\langle R \right| - \left| L \right\rangle \left\langle L \right|$.}
 of the Bloch vector $\mathbf{r}$ with $\rho_P= (\mathbbm{1} + \mathbf{r}\mathbf{\sigma})/2$, which correspond to the Stokes parameters.  
This is done, e.g., by means of  polarization filters that cover the full beam cross section and integrating the total beam intensity after the filters. \TK{ Alternatively, the total beam intensity after the filters can be determined by detecting and integrating the modal weights  $|\ex{b_i}{u_{R,L}}|^2$ (cp. (\ref{pmatrix1})) with respect to any basis set of  spatial modes  $\ket{b_i}$. We discuss the application of such a  modal decomposition technique below.  } 

\begin{figure*}[t]
\centerline{\includegraphics[width=15cm]{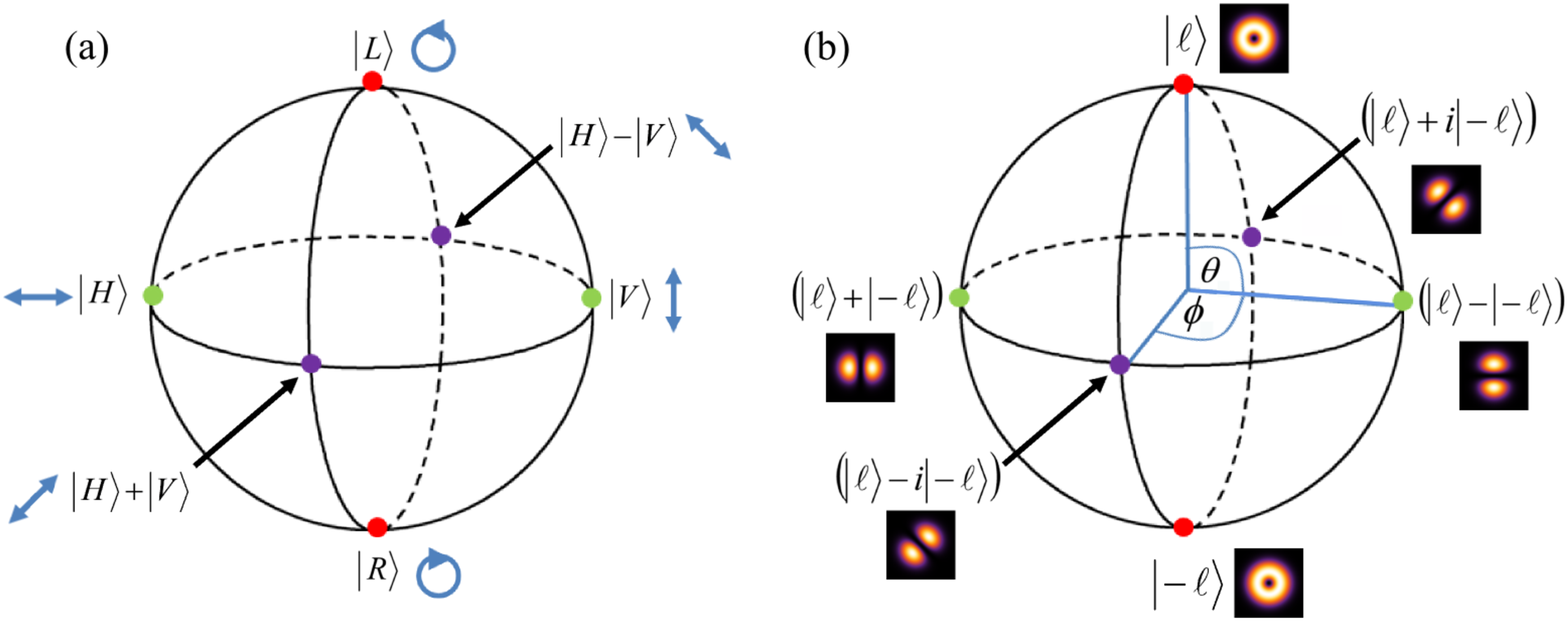}}
\caption{(a) Poincar\'e sphere representation for the states of polarization. (b) An analogous representation for orbital angular momentum states on the Bloch sphere. }
\label{Spheres}
\end{figure*}

Since the reduced density matrix $\rho_P$  (\ref{pmatrix}) describes the state of polarization averaged over a transversal plane, in general it resembles the polarization matrix of an incoherent mixture,  even though the vector beam including the spatial dependence is characterized by a coherent superposition $\ket{\Psi}$. 
This kind of decoherence is also an indicative trait of pure entangled quantum states, where the reduced density matrix representing the state of a particular subsystem or degree of freedom  appears mixed.        
Only in the extreme case of a beam with homogenous polarization $\ket{P}$, i.e. for a scalar beam, the reduced density matrix reflects a pure state, $\rho_P=\left|P \right\rangle \left\langle P \right|$.  
For vector beams, $\rho_P$ can be written as a mixture of pure states: $ \rho_P = \sum_i p_i \left|P_{i} \right\rangle \left\langle P_{i} \right|$. 
The entanglement can thus be quantified in terms of  the "mixedness" of $\rho_P$ as given by the von Neumann entropy (cp. Eq. (\ref{entanglemententropy})). The latter yields the Shannon entropy of the statistical weights $p_i$  of a decomposition of $\rho_P$ with respect to orthonormal states $\ket{P_i}$. This is given by the spectral decomposition of $\rho_P$ , where the spectral values are the weights $p_i$. The  eigenvalues of $\rho_P$ read $(1\pm r)/2$, and  $r=|| \mathbf{r}|| $, the length of the Bloch vector, is given by 
\begin{equation}
r(\rho_P)= (\tr[\rho_P^2])^{1/2} = \left(\sum_i \,\langle\sigma_i\rangle^2\right)^{1/2}.
\label{eq:purity}
\end{equation}
Hence, the entropy of entanglement can be expressed as 
\begin{equation}
E(\ket{\Psi}) = h\left(\frac{1+r}{2}\right),
\label{eq:entropy}
\end{equation} 
where $h(x)=-x\log( x) - (1-x)\log(1-x)$ is the binary entropy.
Please note, that $r$ is a measure of mixedness of $\rho_P$ and thus vectorness of $\rho$ in its own right. Optically it corresponds to the degree of polarization of the averaged polarization state. 

For fluctuating paraxial beams the state of light is characterized by an incoherent mixture rather than a coherent superposition. Therefore, for a general vector beam the beam coherence-polarisation matrix (\ref{BCP}) is given by the ensemble average of the matrix elements in Eq.~(\ref{BCP}) with respect to the different realizations of the electric field $\mathbf{E}$. This includes the time-average of the matrix elements over the "exposure time" of the detector, since the fields $u_R$ and $u_L$ are in general time dependent. For incoherent vector beams with mixed states, the degree of  ``mixedness'' (entropy) of the reduced density matrix can obviously not be directly used to determine the degree of vectorness of the beam. While the notion of entanglement entropy $E$ can be extended to mixed states, it is in general very difficult to calculate. An exception \TK{a vector beam with only two orthonormal spatial modes populated. In this case the spatial degree of freedom and the polarization form a pair of two-level systems and }   $E$ can be determined for all mixed \TK{beam} states by means of  the concurrence $C$:       
\begin{equation}
E(\rho) = h\left(\frac{1+\sqrt{1-C^2}}{2}\right)\,.
\end{equation}
Because of the one-to-one correspondence between $C$ and $E$, the concurrence represents an alternative entanglement measure for pairs of two-level systems. \cite{Wootters2001}. It is  given by 
\begin{equation}
\mathcal{C}(\rho) = \max\{0,\sqrt{\lambda_1} - \sqrt{\lambda_2} - \sqrt{\lambda_3} - \sqrt{\lambda_4}\} ,
\end{equation}
with $\lambda_i$ being the eigenvalues in decreasing order of the Hermitian matrix
\begin{equation}
R = \rho(\sigma_y\otimes\sigma_y)\rho^*(\sigma_y\otimes\sigma_y) ,
\label{eqn:R}
\end{equation}
where $^*$ represents the complex conjugate and $\sigma_y$ is the Pauli $y$-matrix
\begin{equation}
\sigma_y =  \left[ \begin{array}{ll} 0 & -{\rm i} \\ {\rm i} & 0 \end{array} \right] .
\end{equation}
In order to use the concurrence to determine the degree of vectorness of a paraxial beam, the spatial degree of freedom has to be projected onto a two-dimensional subspace. This could for example be the space spanned by two Hermite Gaussian or Laguerre Gaussian modes and can be accomplished by using mode filters.  In this regard we point out that, analogously to polarisation states on the Poincar\'e sphere, we can depict OAM states of a reduced subspace on an equivalent sphere\cite{Padgett1999}. For example, the left- and right-handed helicities of the OAM states that lie on the poles of this sphere, correspond the left-and right-handed circularly polarized states on the Poincar\'e sphere (see Fig.~\ref{Spheres}). Afterward this projection of the spatial degree of freedom, the elements of the resulting $4\times4$ beam coherence-polarisation matrix can be detected by means of state tomography.

In the following we demonstrate experimental methods for both -- to measure the degree of vectorness by means of state tomography and by detecting the degree of (averaged) polarization. Moreover, we introduce a third method that is based on measuring the violation of a Bell inequality.


\begin{figure*}[t]
\centerline{\includegraphics[width=18cm]{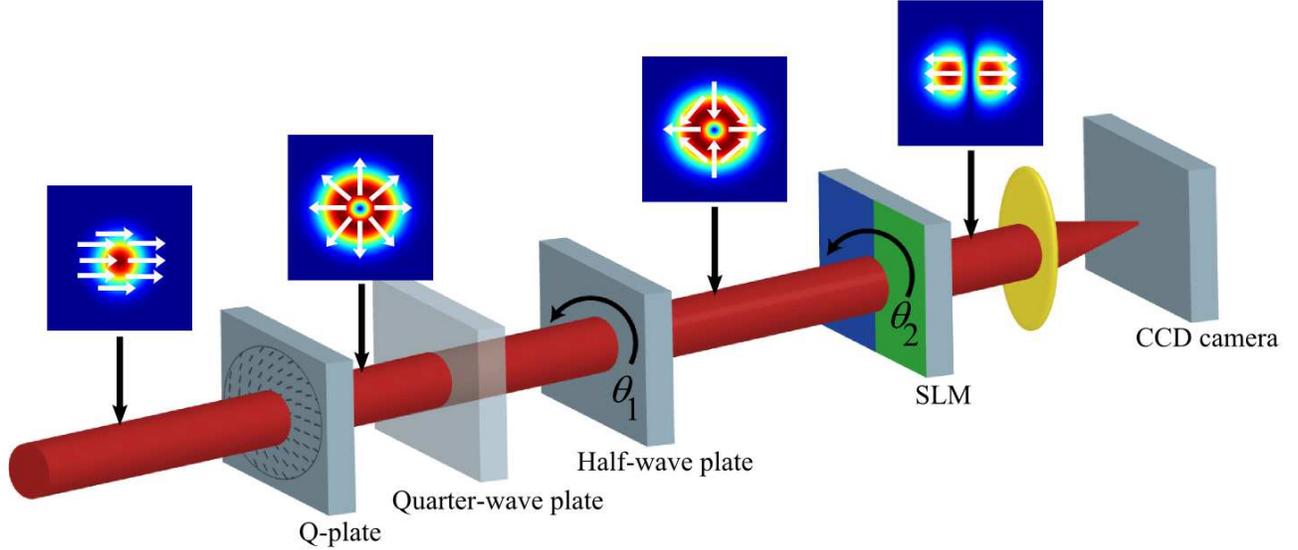}}
\caption{Schematic representing the experimental procedure with which to generate and measure cylindrical vector (CV) beams. A horizontally polarized Gaussian beam incident on the q-wave plate is converted to a radially polarized vortex beam. The q-plate converts left-handed circularly (LHC) polarized light into right-handed circularly (RHC) polarized light and adds OAM of +1 to the incident beam, while RHC is converted to LHC and OAM of -1 is added to the incident beam. Thus, the generated beam is a superposition of OAM modes, $\left|\ell = 1 \right\rangle + \left|\ell = -1\right\rangle$, with radially-varying polarization. A quarter-wave plate was only inserted to measure the circularly polarized modes and was set at an orientation of $\pi/4$~radians. The half-wave plate is then rotated by an angle $\theta_{1}$, thereby measuring the polarization of the beam. The hologram encoded onto the SLM represents a superposition of OAM modes, $\left| \ell = 1\right\rangle + \exp(i\theta_{2})\left| \ell = -1\right\rangle$, where $\theta_{2}$ is the orientation of the hologram.  The SLM, Fourier lens and CCD camera together form the OAM-polarization detection component using a modal decomposition technique. The SLM only directs horizontally polarized light into the first diffraction order and therefore simultaneously acts as a polarizer while also performing an azimuthal decomposition.}
\label{Setup}
\end{figure*}
Our experimental setup can be seen in Fig.~\ref{Setup}, and can be divided into two parts: the preparation of the vector beam and the measurement of its non-separability.  Without any loss of generality we have chosen to create vector vortex modes that lie on the higher-order Poincare sphere\cite{Milione2011}.
We made use of a q-plate \cite{Marrucci2006} to prepare vector vortex modes. The q-plate uses a spatially variant geometric phase to couple polarization to orbital angular momentum following the selection rules:
\begin{eqnarray}
\left| \ell, L \right\rangle \rightarrow \left| \ell + Q, R \right\rangle, \\
\left| \ell, R \right\rangle \rightarrow \left| \ell - Q, L \right\rangle.
\label{eq:Qplate}
\end{eqnarray}
The azimuthal charge introduced by the q-plate is $Q = 2q$. The polarization distribution after the q-plate depends on the initial incident polarization state. That is, horizontally polarized light will be transformed into a radially-varying polarization state, while an azimuthally-varying state is created from vertically polarized light incident on the q-plate. In our experiment $q = 1/2$ and the incident Gaussian beam was horizontally polarized with $\ell = 0$, thereby generating a radially polarized vortex mode consisting of a superposition of $\ell = \pm 1$ modes. 
Thus, the field after the half-wave plate can be described by
\begin{equation}
\left| \Psi \right\rangle = \sqrt{a}\left| \ell = 1\right\rangle \left| R \right\rangle \pm \sqrt{1-a}\left| \ell = -1\right\rangle \left| L \right\rangle,
\label{equ:wave}
\end{equation}
where $\left| \ell = \pm 1\right\rangle$ represent the azimuthal components of the vortex beam.   

To perform the necessary projections and state tomography we require only a waveplate and \TK{a} spatial  light modulator (SLM). In an entanglement setup, separate projections of the same property are performed simultaneously on a pair of entangled photons. For example, the polarization states are measured by placing a polarizer in the path of each photon and the simultaneous arrival of the photons is recorded. Analogously, the OAM of each photon can be measured using SLMs as mode-specific filters. In the case of vector beams, both degrees of freedom must be measured locally: polarization and OAM.  The measurement of states on the OAM subspace is performed by modal decomposition\cite{Litvin2012} using an SLM. By considering the modal decomposition of the input field $u$ into the azimuthal modes $\exp(i\ell\phi)$ such that $u = \sum_{\ell} a_{\ell} \exp(i\ell \phi)$,  \TK{ the modulus} of the modal  weighting coefficients $a_{\ell}$ can be determined by the inner product of the field with an azimuthal match filter. That is, \TK{$\left| \left\langle u | \exp(i\ell\phi)\right\rangle\right| = |a_{\ell}|$,} which can be experimentally performed by directing the field $u$ onto an SLM encoded with the match filter hologram, $\exp(-i\ell\phi)$, and recording the on-axis intensity on a CCD camera after Fourier lens $L_{1}$ in Fig.~\ref{Setup}. In this experiment, we generated vortex modes carrying a superposition of OAM values $\ell = \pm 1$. 

SLMs are also polarization sensitive in that the desired beam reflected from the screen consists of only horizontally polarized light. As such, the SLM acts as a horizontal polarizer, which, when rotated, acts as a filter for the linear polarization states: horizontal, vertical, diagonal and anti-diagonal. As a matter of practicality, we fixed the SLM to reflect only horizontally polarized light and instead inserted a half-wave plate before the SLM, which we rotate to realise a filter for the linear polarisation states. By inserting an additional element, a quarter-wave plate, we were able to filter the circularly polarized components. The quarter-wave and rotation of the half-wave plate performing all the necessary projective measurements as outlined in the theory.

We first performed a Bell-type inequality measurement to demonstrate a violation of Bell's inequality using vector vortex modes. Typically, a Bell inequality is performed on an entangled pair of photons and using a single degree of freedom, e.g., polarization \cite{Aspect1981} or OAM \cite{Leach2009}. In our experiment, instead of measuring one degree of freedom non-locally (e.g., two separated photons) we measure two degrees of freedom locally, i.e., on the same classical field. The Bell parameter $S$ satisfies the inequality $-2 \leq S \leq 2$ for classical correlations, in the case of entanglement, or for scalar beams, in the case of classical fields. 
We define the Bell parameter $S$ to be,
\begin{equation}
S = E(\theta_{1}, \theta_{2}) -  E(\theta_{1}, \theta_{2}^{\prime}) +  E(\theta_{1}^{\prime}, \theta_{2}) +  E(\theta_{1}^{\prime}, \theta_{2}^{\prime}),
\label{eq:Bell}
\end{equation}
where $E(\theta_{1}, \theta_{2})$ can be calculated by measuring the on-axis intensity $I(\theta_{1}, \theta_{2})$ on the camera:

\begin{widetext}
\begin{equation}
E(\theta_{1}, \theta_{2}) = \frac{I(\theta_{1}, \theta_{2}) + I(\theta_{1}+\frac{\pi}{2}, \theta_{2}+\frac{\pi}{2}) - I(\theta_{1}+\frac{\pi}{2}, \theta_{2}) - I(\theta_{1}, \theta_{2}+\frac{\pi}{2})}{I(\theta_{1}, \theta_{2}) + I(\theta_{1}+\frac{\pi}{2}, \theta_{2}+\frac{\pi}{2}) + I(\theta_{1}+\frac{\pi}{2}, \theta_{2}) + I(\theta_{1}, \theta_{2}+\frac{\pi}{2})}.
\label{eq:Eparameter}
\end{equation}
\end{widetext}

\begin{figure}[t]
\centerline{\includegraphics[width=8.3cm]{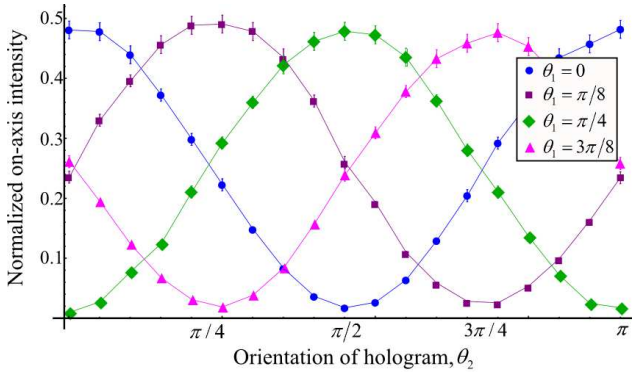}}
\caption{Bell-type curves for four different orientations of the half-wave plate: $\theta_{1} = 0$, $\theta_{1} = \pi/8$, $\theta_{1} = \pi/4$ and $\theta_{1} = 3\pi/8$. For each change in polarization distribution, the hologram representing the superposition state $\left|\ell = 1 \right\rangle + \exp(i\theta_{2})\left|\ell = -1 \right\rangle$ was rotated by $\theta_{2} = \{0, \pi\}$ radians and the on-axis intensity was recorded.}
\label{BellCV}
\end{figure}

Here $\theta_{1}$ and $\theta_{2}$ are the angles of orientation of the half-wave plate and the encoded hologram, respectively. For each orientation of the wave plate, the holograms were rotated from $\theta_{2} = 0$ to $\theta_{2} = \pi$ and the on-axis intensity was recorded. This was repeated for four different orientations of the half-wave plate: $\theta_{1} = 0$ rad, $\theta_{1} = \pi/8$ rad, $\theta_{1} = \pi/4$ rad and $\theta_{1} = 3\pi/8$ rad, as shown in Fig.~\ref{BellCV}.   

From Eq.~\ref{eq:Bell} and \ref{eq:Eparameter} we found our Bell parameter to be $S = 2.72 \pm 0.02$. We have demonstrated a violation of Bell's inequality by 36 standard deviations for these vector vortex modes. This highlights the non-separability of the classical mode. 

\begin{figure}[t]
\centerline{\includegraphics[width=8.3cm]{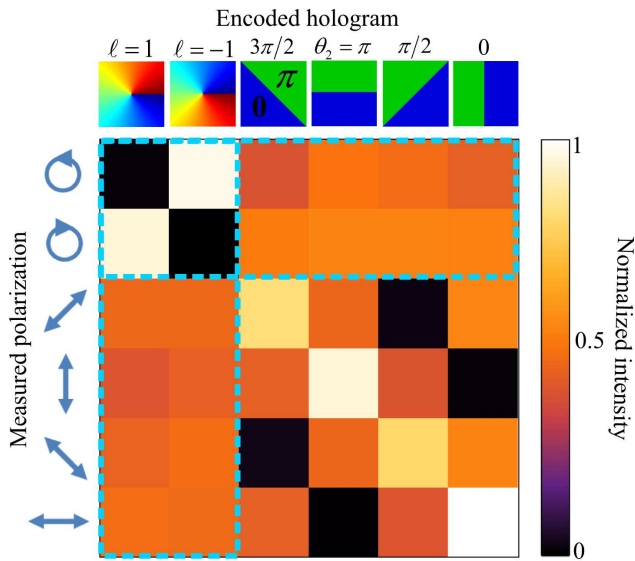}}
\caption{Experimental measurements obtained from a full state tomography measurement. The polarization was measured using a combination of a quarter-wave plate and half-wave plate. For each polarization state, the field was decomposed in the $\ell = |1|$ OAM basis, which includes both pure and superposition states, and the on-axis intensity was recorded. The measurements outlined in blue were used to calculate the entanglement entropy of the vector beam, where either the pure OAM states (first two columns) or the circularly polarized states (first two rows) were chosen as the basis state.}
\label{Tomography}
\end{figure}

Next we performed a full state tomography measurement \cite{Jack2009, Agnew2011} to calculate the density matrix of the state. In this measurement we required not only the superposition states of polarization and OAM but also the pure states: left and right circular polarization and $\ell = \pm 1$ OAM modes. In terms of the higher-order Poincar\'e sphere, the pure modes are represented at the poles of the sphere \cite{Milione2011}. For each of the six polarization states (right, left, horizontal, diagonal, vertical and anti-diagonal), a modal decomposition was executed using six different holograms: $\left|\ell = 1 \right\rangle$, $\left|\ell = -1 \right\rangle$, $\left|\ell = 1 \right\rangle + \exp(i\theta_{2})\left|\ell = -1 \right\rangle$ for $\theta_{2} = 0, \pi/2, \pi, 3\pi/2$. Figure~\ref{Tomography} shows the normalized intensity measurements for each of the six polarization states and the six OAM states.
This tomographic method produces an over-complete set of 36 measurements, which can be used to minimize the Chi-square quantity and reconstruct the density matrix $\rho$. The degree of vectorness of any field can then be calculated from the density matrix. We found the concurrence, a measure of the degree of vectorness, of our vectorial field to be $C = 0.96 \pm 0.01$, while that of a scalar state to be $C = 0.09 \pm 0.01$. A maximal vector field is represented by 1, while a value of 0 represents a purely scalar field.

\begin{table}[ht] 
\caption{Comparison between a vector and scalar beam for three different measurements: a Bell inequality measurement, a concurrence measurement and an entropy measurement. A Bell parameter of greater than 2 represents a vector beam, while a value close to 1 for both the concurrence and entropy represents a vector beam. The errors for the scalar beam were insignificant.} 
\centering 
\begin{tabular}{c c c c} 
\hline\hline 
 & Bell parameter & Concurrence & Entropy \\ [1ex] 
\hline 
Vector & 2.72 $\pm$ 0.02 & 0.96 $\pm$ 0.01 & 0.98 $\pm$ 0.01 \\ 
Scalar & 0.10  & 0.09  & 0.02  \\  [1ex] 
\hline 
\end{tabular} 
\label{table:results} 
\end{table} 

In fact, two states selected on the polarisation (or OAM) sphere and projective measurements on the OAM (or polarisation) is all that is required for the measurement of the entanglement entropy. This space within the full tomography measurements is shown within the dashed lines of Fig.~\ref{Tomography}.  The  purity $r$ of the reduced density matrix in Eq.~\ref{eq:purity} was calculated by first selecting a \TK{spatial basis}, e.g. the pure OAM states ($\left| \ell = \pm 1\right\rangle$) shown as the first two columns in Fig.~\ref{Tomography}, and then calculating \TK{ the components of the Bloch vector. For example, $\langle\sigma_3\rangle = \tr[(\left|R \right\rangle \left\langle R \right|- \left|L \right\rangle \left\langle L \right|)\rho_p] = \sum_{i=-1,1} |\ex{l=i}{u_R}|^2 - |\ex{l=i}{u_L}|^2  $.} Similarly for $\sigma_{1}$ and $\sigma_{2}$, where the horizontal/vertical states and the diagonal/anti-diagonal states are considered, respectively. This resembles a degree of polarization measurement and of course, any \TK{spatial basis}  can be chosen. However, the Poincar\'e and Bloch spheres in Fig.~\ref{Spheres} clearly illustrate the analogy between polarization and an OAM subspace and as such, we can also \TK{ first }choose a polarization basis, e.g. the circularly polarized states shown as the first two rows in Fig.~\ref{Tomography}, to calculate the purity $r$. Using Eq.~(\ref{eq:entropy}), we calculated the entanglement entropy for the vector vortex beam to be $0.98 \pm 0.01$, which indicates a high level of non-separability or vectorness. Table~\ref{table:results} compares the results of the three techniques for a vector and scalar beam.  We find that all the tools provide an accurate measure of the classical field, and that in particular we are able to quantify the degree of vectorness of the field.  In the 1990s statistical tools were applied to laser beam characterisation and led to the now ubiquitous beam quality factor ($M^2$) as a measure of modal content and divergence; here we provide a new measure for the vector nature of classical fields through the use of quantum tools.

Using measurement techniques more commonly associated with quantum entanglement, we have demonstrated methods to distinguish between scalar and vector beams and to determine the degree of non-separability, or vectorness, of the vector beam. It should be noted that this is a very practical tool - to date no single measure exists for the vectorness of such fields.  The suitability of the quantum tools to this problem lies in the common property that vector beams share with entangled states: non-separability. In some sense these fields may be considered akin to entangled states, and hence the phrase ``classical entanglement'' has emerged.  In this work we have shown that indeed the measurement tools of entanglement are highly appropriate to such classically non-separable fields.  One wonders then: if the quantum tools are so applicable to these classically entangled fields, are the classically entangled field then applicable as vehicles to realise quantum processes that rely on entanglement?

\section{Appendix}

\subsection{Action of half-wave plate}
\label{Halfwave}
The spatial light modulator (SLM) only acts on the horizontally polarized component of a beam. As such, if the SLM is rotated, it will act on the different states of polarization. The rotation of the half-wave plate in the experimental setup is equivalent to rotating the SLM, as the half-wave plate changes the polarization state of the beam and the SLM, thereafter, acts on the horizontal component. A half-wave plate can be written as a Jones matrix as follows,
\begin{equation}
\left( \begin{array}{cc}
\cos(2\theta) & \sin(2\theta)  \\
\sin(2\theta)  & -\cos(2\theta)
 \end{array} \right) ,
 \label{eq:half}
\end{equation}
where $\theta$ is the angle between the fast axis and the horizontal axis. By setting $\theta = 0$~radians, the polarization of a field $u$ is transformed as:
\begin{equation}
\left( \begin{array}{c}
u_{x}  \\
u_{y}  
 \end{array} \right)
\xrightarrow{0~\textrm{rad}}
 \left( \begin{array}{c}
u_{x}  \\
-u_{y}  
 \end{array} \right) .
 \label{eq:Zero}
\end{equation}
As the SLM only acts on the horizontal component of the field, we see that we measure the original horizontal component. However, if the plate is rotated such that $\theta = \pi/4$~rad, then we measure the vertical component of the original beam.
\begin{equation}
\left( \begin{array}{c}
u_{x}  \\
u_{y}  
 \end{array} \right)
\xrightarrow{\pi/4~\textrm{rad}}
 \left( \begin{array}{c}
u_{y}  \\
u_{x}  
 \end{array} \right) .
 \label{eq:45}
\end{equation}
Similarly, the diagonal and anti-diagonal polarization components can be measured by orientating the plate at $\theta = \pi/8$ and $\theta = 3\pi/8$~rad, respectively. This is seen in Eq.~(\ref{eq:22}) and (\ref{eq:67}).
\begin{equation}
\left( \begin{array}{c}
u_{x}  \\
u_{y}  
 \end{array} \right)
\xrightarrow{\pi/8~\textrm{rad}}
 \frac{1}{\sqrt{2}}\left( \begin{array}{c}
u_{x} + u_{y}  \\
u_{x} - u_{y} 
 \end{array} \right) ,
 \label{eq:22}
\end{equation}
and
\begin{equation}
\left( \begin{array}{c}
u_{x}  \\
u_{y}  
 \end{array} \right)
\xrightarrow{3\pi/8~\textrm{rad}}
 \frac{1}{\sqrt{2}}\left( \begin{array}{c}
-u_{x} + u_{y}  \\
u_{x} + u_{y} 
 \end{array} \right). 
 \label{eq:67}
\end{equation}
Thus, the half-wave plate together with the polarization-sensitive SLM, each polarization state along the equator of the Poincar\'e sphere can be measured.

\subsection{Action of quarter-wave plate}
A quarter-wave plate was inserted between the q-plate and half-wave plate in Fig.~\ref{Setup} to measure the circular polarization states of the vector beam. The Jones matrix for a quarter-wave plate is written as:
\begin{equation}
\left( \begin{array}{cc}
\cos^{2}\theta + i \sin^{2}\theta & \sin\theta\cos\theta - i\sin\theta\cos\theta  \\
\sin\theta\cos\theta - i\sin\theta\cos\theta  & \sin^{2}\theta + i \cos^{2}\theta
 \end{array} \right) .
 \label{eq:quarter}
\end{equation}
Again, $\theta$ is the angle between the fast axis of the plate and the horizontal axis. Setting the plate at an orientation with $\theta = \pi/4$, the Jones matrix becomes:
\begin{equation}
\frac{\exp(i\pi/4)}{\sqrt{2}}\left( \begin{array}{cc}
1 & -i  \\
-i  & 1
 \end{array} \right) .
 \label{eq:quarter45}
\end{equation}
Thus, the action of a quarter-wave plate orientated at $\theta= \pi/4$ on a field $u$ is described by
\begin{equation}
\left( \begin{array}{c}
u_{x}  \\
u_{y}  
 \end{array} \right)
\xrightarrow{\pi/4~\textrm{rad}}
 \frac{\exp(i\pi/4)}{\sqrt{2}}\left( \begin{array}{c}
u_{x} - iu_{y}  \\
-u_{x} + u_{y} 
 \end{array} \right). 
 \label{eq:q45}
\end{equation}
Thus, looking at the horizontal component only, the quarter-wave plate allows us to measure right-circularly polarized light. Followed by the half-wave plate oriented at either $\theta = 0$ or $\theta = \pi/4$, we can measure right-and left- circularly polarized light, respectively.

\end{document}